\documentclass[%
 reprint,
 twocolumn,
superscriptaddress,
showpacs,
showkeys,
 amsmath,amssymb,
 prd,
]{revtex4}

\usepackage{color}
\usepackage{lipsum}

\usepackage{hyperref}
\usepackage{theorem,algorithm,algorithmic}
\usepackage{footnote}
\makesavenoteenv{table}
\usepackage{graphicx}
\usepackage{dcolumn}
\usepackage{bm}
\usepackage{soul} 

\DeclareMathOperator\erf{erf}

\begin{document}

\preprint{APS/123-QED}

\title{Dealing with missing data in the MICROSCOPE space mission: \\An adaptation of {\it inpainting} to handle colored-noise data.}

\author{Sandrine Pires}
 \email{sandrine.pires@cea.fr}
\affiliation{
 Laboratoire AIM, CEA/DSM--CNRS 
 Universit\'e Paris Diderot 
  IRFU/SAp-SEDI, CEA Saclay 
   Orme des Merisiers 
    91191 Gif-sur-Yvette, France
}%
\author{Joel Berg\'e}
\affiliation{ONERA - The French Aerospace Lab, 29 avenue de la Division Leclerc, 92320 Ch\^atillon, France}

 \author{Quentin Baghi}
 \affiliation{ONERA - The French Aerospace Lab, 29 avenue de la Division Leclerc, 92320 Ch\^atillon, France}

 \author{Pierre Touboul}
 \affiliation{ONERA - The French Aerospace Lab, 29 avenue de la Division Leclerc, 92320 Ch\^atillon, France}
 
  \author{Gilles M\'etris}
  \affiliation{Geoazur (UMR 7329), Observatoire de la C\^ote d'Azur Bt 4, 250 rue Albert Einstein, Les Lucioles 1, Sophia Antipolis, 06560 Valbonne, France}

\date{\today}

\begin{abstract}


The MICROSCOPE space mission, launched on April 25, 2016, aims to test the weak equivalence principle (WEP) with a $10^{-15}$ precision. To reach this performance requires an accurate and robust data analysis method, especially since the possible WEP violation signal will be dominated by a strongly colored noise. An important complication is brought by the fact that some values will be missing --therefore, the measured time series will not be strictly regularly sampled. Those missing values induce a spectral leakage that significantly increases the noise in Fourier space, where the WEP violation  signal is looked for, thereby complicating scientific returns.
Recently, we developed an {\it inpainting} algorithm to correct the MICROSCOPE data for missing values.
This code has been integrated in the official MICROSCOPE data processing and analysis pipeline because it enables us to significantly measure an equivalence principle violation (EPV) signal in a model-independent way, in the inertial satellite configuration.
In this work, we present several improvements to the method that may allow us now to reach the MICROSCOPE requirements for both inertial and spin satellite configurations. The main improvement has been obtained using a prior on the power spectrum of the colored-noise that can be directly derived from the incomplete data. 
We show that after reconstructing missing values with this new algorithm, a least-squares fit may allow us to significantly measure an EPV signal with a $0.96\times10^{-15}$ precision in the inertial mode and $1.20\times10^{-15}$ precision in the spin mode.
Although, the {\it inpainting} method presented in this paper has been optimized to the MICROSCOPE data, it remains sufficiently general to be used in the general context of missing data in time series dominated by an unknown colored-noise. The improved {\it inpainting} software, called ICON ({\bf I}npainting for {\bf CO}lored-{\bf N}oise dominated signals), is freely available at \url{http://www.cosmostat.org/software/icon}.


\end{abstract}

\maketitle


\section{Introduction} 
\label{intro}
The MICROSCOPE space mission aims to test the weak equivalence principle (WEP) with a precision of $10^{-15}$. This is more than two orders of magnitude better than the current ground-based constraints \cite{EPV:will14}, and will allow a precise test of general relativity after LIGO brought a new confirmation of its predictions by the direct detection of gravitational waves \cite{ligo:abbott16}.
Indeed, the equivalence principle is a cornerstone of the general theory of relativity. It states in particular that the inertial mass and the gravitational mass are equivalent; in other words, the acceleration imparted to a body by a gravitational field is independent of its mass and its composition.
Any detection of a WEP violation would be paramount since it may question the very basis of general relativity and, more generally, our understanding of the Universe.
On the opposite, confirming the WEP to a precision of $10^{-15}$ would place new constraints on unified models for fundamental interactions, since some predict a violation below $10^{-13}$ \cite{wep:damour12, wep:overduin12}.

Estimating a possible equivalence principle violation (EPV), or lack thereof, will require a careful data analysis, and especially a fine characterization of the measurement noise. A possible difficulty, identified in \cite{karma:baghi15, inpainting:berge15}, originates from missing data in the time series resulting in irregularly sampled data. Although missing data are a common problem in physics experiments (they occur when data acquisition fails or when a contamination from an external source invalidates some points), they are particularly troublesome when the measurement noise is colored (i.e., frequency-dependent). 
In particular,  \cite{karma:baghi15, inpainting:berge15} have shown that the presence of gaps has a strong impact on the precision of the ordinary least squares fits of harmonic signals. 
This is due to the spectral leakage of the noise in the frequency domain, which increases the uncertainty of the fit by several orders of magnitude, even for a small fraction of missing data.

An important effort has been made to reduce the impact of missing values in the MICROSCOPE data analysis. 
A Kalman-Auto-Regressive Model Analysis (KARMA) has been proposed in \cite{karma:baghi15}, which generalizes least-squares estimation to missing data problems without ``filling'' missing values. 
As it has been shown that KARMA allows us to reach MICROSCOPE's requirements, it has been integrated in the official MICROSCOPE data analysis pipeline. 
However, to strengthen our conclusions after analyzing the MICROSCOPE data, and since missing data are a crucial difficulty in the data analysis, we want to have at least two independent techniques to deal with them. That is why we have proposed an alternative route in \cite{inpainting:berge15}: 
we use the {\it inpainting} algorithm to fill in data gaps, which then allows us to use an ordinary least-squares method to look for and characterize a possible EPV. We showed that this method allows us to reach also the MICROSCOPE requirements in the so-called ``inertial'' mode, where the MICROSCOPE spacecraft is kept fixed with respect to distant stars. However, this method fails in the ``spin'' mode, where the spacecraft rotates about the axis normal to the orbital plane. Hence, the method needed to be improved so that it can be used in both experimental modes. This is the aim of this paper, which can be seen as the natural sequel to \cite{inpainting:berge15} (thereafter Paper I).

Paramount to MICROSCOPE, either to characterize the confidence level of an EPV detection, or to characterize a new upper limit on the WEP (if no EPV is detected), is the noise characterization. This characterization will be done in Fourier space, and therefore amounts to estimating the noise power spectral density (PSD). As explained above, the PSD estimation is affected by the spectral leakage created by missing data. Methods like KARMA and {\it inpainting} (Paper I), although they allow us to reach MICROSCOPE's requirements in the characterization of a possible EPV detection, fail to fully estimate the PSD \cite{karma:baghi15, inpainting:berge15}, and are therefore not fully satisfactory. In \cite{karma2:baghi16}, the authors extended the KARMA technique into a modified expectation-conditional-maximization (M-ECM) technique; using simulated data, they showed that this technique allows us to fully estimate the MICROSCOPE's noise PSD. 
This new paper aims to provide a description of the several improvements that have been brought to the previously developed {\it inpainting} method (Paper I) in the search of reconstructing the full noise PSD. Those improvements enable us to reach the MICROSCOPE's scientific goals in both inertial and spin modes.

This paper is organized as follows. 
In Sec.~\ref{sect_microscope}, we summarize the MICROSCOPE mission and review how missing data affect the data analysis.
Sec.~\ref{sect_inpainting} describes the improvements brought to the {\it inpainting} method and their motivations. The major improvement has been obtained by adding a prior on the noise power spectrum directly derived from the data. The results are presented in Sec. \ref{sect_results} based on MICROSCOPE mock data; in particular, we show the gain in precision that we obtain in the evaluation of a possible EPV signal with a Least Square fit compared to the previous method. 
We conclude in Sec.~\ref{sect_conclusion}.

\section{The MICROSCOPE mission} 
\label{sect_microscope}


MICROSCOPE will test the WEP by measuring the relative acceleration of two test masses of different composition freely falling in the earth's gravitational field. To achieve the highest possible stability and accuracy, the test masses are on-board a drag-compensated and attitude-controlled satellite which screens them from non-gravitational accelerations. The science data will consist of time series of differential accelerations (the half-difference of the test masses accelerations) measured along a sensitive axis by onboard inertial sensors.
The source of the gravitational signal used to test the WEP is the earth�s gravitational field modulated by the motion of the satellite as it orbits Earth: hence, the EPV signal is a sine expected at a well known frequency $f_{\rm EP}$ (which depends on the satellite�s motion) along the most sensitive axial axis of the instrument."

The measured signal in the MICROSCOPE experiment can thus be written as:
\begin{equation}
X_{\rm EP}(t) = \frac{1}{2} \mathcal{M}_{\rm EP} \delta g_{\rm EP}(t) + \mathcal{S}(t) + \mathcal{N}(t),
\end{equation}
where $\mathcal{M}_{\rm EP}$ is an instrumental calibration factor \cite{microscope:touboul09}, $\delta$ is the EPV parameter we aim to detect and characterize, $g_{\rm EP}(t)$ is the Earth gravity field's projection on the measurement-axis (with $g\approx 8$ m.s$^{-2}$ at MICROSCOPE's altitude --700 km) \cite{proceeding:berge15}, $\mathcal{S}$ represents systematics errors, and $\mathcal{N}$ is the statistical inertial sensor noise.

Assuming perfect correction of systematics $\mathcal{S}$ and instrument's calibration $\mathcal{M}_{\rm EP}$ (these corrections are out of the scope of this paper), we will measure a sine-wave signal ($\delta g_{\rm EP} (t)/2$) at frequency $f_{\rm EP}$ dominated by a colored-noise ($\mathcal{N} (t)$).

Different experimental modes can be used, which will allow us to confirm (or exclude) an EPV detection. In particular, the inertial mode is defined as keeping the satellite's attitude fixed with respect to distant stars; in this case, the WEP test frequency is equal to the orbital frequency $f_{\rm EP,~iner} = f_{\rm orb} = 1.8 \times 10^{-4}$ Hz. In the spin mode, the satellite rotates about the axis normal to the orbital plane in opposite to the rotation due to to the orbital motion, with a frequency $f_{\rm spin}$, thereby offset the WEP test frequency to $f_{\rm EP, spin} = f_{\rm orb} + f_{\rm spin}$; by choosing $f_{\rm spin}>0$, we increase $f_{\rm EP}$, moving it to a frequency where the measurement noise is lower. In this paper, we assume $f_{\rm spin} = 4.5 f_{\rm orb}$, thereby $f_{\rm EP,~spin} = 10^{-3}$Hz. The orbital frequency can be measured in-flight with precise orbit determination; it is then used in the data analysis to look for the EPV signal through a Least-Square fit.

The duration of each session is chosen in such a way to ensure a measurement noise of about $4 \times 10^{-15}$ ms$^{-2}$ on the differential acceleration at $f_{\rm EP}$, as needed to reach a $10^{-15}$ precision on the EPV parameter $\delta$. Thus, the inertial and spin sessions last respectively 120 and 20 orbits (inertial and spin sessions will be performed sequentially, one after another). The MICROSCOPE data analysis challenge is then to be able to detect and estimate the amplitude of the periodic signal with a precision of $4 \times 10^{-15}$ ms$^{-2}$ in both the inertial and spin modes. In case no violation is detected, the challenge is to characterize the noise with the same precision.

\begin{figure}
\includegraphics[width=0.45\textwidth]{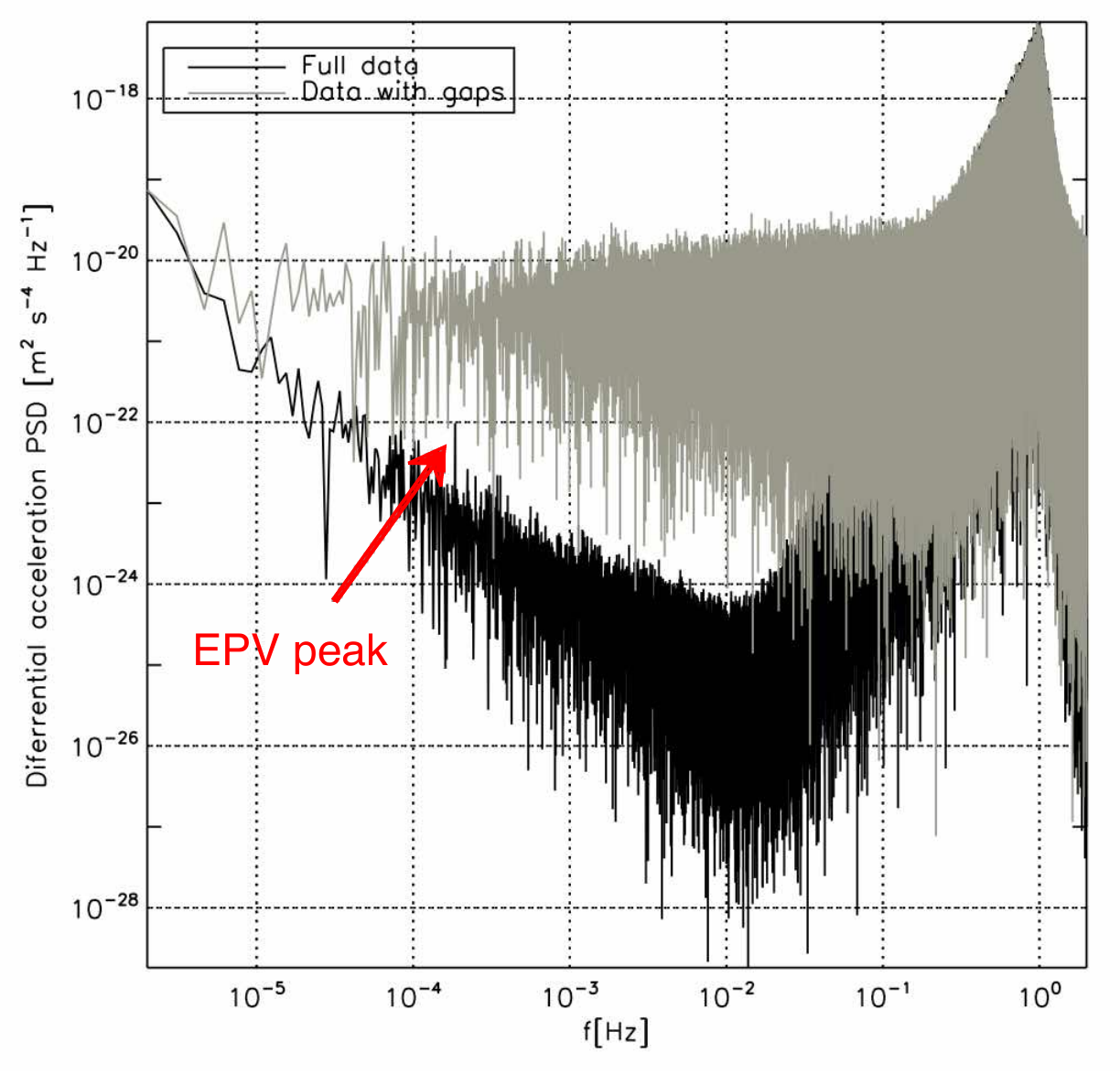}
\caption{The black curve shows the MICROSCOPE PSD estimate for a 120 orbits simulation. An example of a possible EPV signal of $3 \times 10^{-15}$ in the inertial mode is shown by the peak at $1.8\times10^{-4}$ Hz. The grey curve shows the spectral leakage affecting the PSD estimate when gaps are present in the data.}
\label{fig_microscope}       
\end{figure}

Touboul \cite{microscope:touboul09} computed the expected MICROSCOPE's error budget. We use this model to specify the noise PSD (Fig.~\ref{fig_microscope}, black curve). 
In-flight performance sessions will be dedicated to fully and finely characterize the actual noise. We will then be able to take into account any evolution of the noise characteristics in the data analysis. However, we expect the actual in-flight noise PSD to not differ much from the model \cite{microscope:touboul09}, so the results of this paper are robust enough for the upcoming MICROSCOPE data analysis.

Nominally, the measured accelerations are regularly spaced, with a time sampling of 0.25 second. However, missing data are almost inevitable in long observations. Most expected data alteration in MICROSCOPE come from tank crackles and Multi-Layer Insulation (MLI) coating crackles and are shorter than one second.; micrometeorite impacts are expected to be rare and create very short gaps; tele-transmission losses are wider (up to several seconds) but very rare. 
If both accelerometers of a differential accelerometer were perfect, such tank crackes and other spacecraft-related disturbance (such as micrometeorit impacts) would not appear in the differential acceleration, since both accelerometer would measure them in the same way. However, the accelerometers are not perfectly identical, and small differences in their electronics's transfer function create small differences in their measurement. Those differences are compensated for in steady-state regime by a posteriori data analysis. However, they currently cannot be easily compensated for in the transient regime created by glitches; an empirical model from the data should allow us to eventually correct for those glitches without the need to mask them.

In the case of MICROSCOPE, we expect in the worst-case scenario up to about 3\% of missing data (resp. 4\% of missing data) in the inertial mode (resp. in the spin mode). As mentioned in \cite{microscope:hardy13a}, some crackles come from the temperature varaitions of the satellite's multi-layer insulation cover due to changes in the orientation of the satellite with respect to the Earth (the earth's albedo heats one or another side of the satellite along its orbit); in the spin mode, such changes are more frequent, resulting in more crackles, and therefore more missing data than in the inertial mode.
Further discussion about the sources of missing values in MICROSCOPE can be found in \cite{karma:baghi15, microscope:hardy13a}.
Detailing how invalid data points are detected, and how gaps' location and size are set up after the detection such invalid data, goes beyond the scope of this paper. In a few words, we can detect invalid data with a $\sigma$-clipping technique; the size of gaps will be set empirically, to remove any transient behavior after crakles (Berg\'e et al in prep). Rather, we assume that that all such invalid data are correctly detected and masked. We follow \cite{inpainting:berge15} to define the size and distribution of gaps.

Although the fraction of missing data is very small, the detection and estimation of a possible EPV signal is seriously complicated by this data losses and/or alteration.
This is because the spectral leakage induced by missing data makes the noise power from high-power regions spread over the frequency domain; hence, the noise in Fourier space where the EPV signal is looked for is significantly increased, since the measurement noise is strongly colored.
Fig.~\ref{fig_microscope} shows the effect of missing data in the MICROSCOPE PSD estimate (grey curve). Missing data create an important spectral leakage from $f \approx 1$ Hz to surrounding frequencies. As a result, the noise in the band $[10^{-4}-10^{-1}]$Hz, where the EPV signal is looked for, is largely dominated by the spectral leakage from the high-frequency noise. We added to the PSD shown in Fig.~\ref{fig_microscope} an EPV signal of $3\times10^{-15}$ (red arrow); it is clear that the spectral leakage due to missing data makes its detection extremely difficult in Fourier space. On the opposite, without missing data, the EPV peak emerges clearly from the noise (black curve).

\section{Sparse inpainting} 
\label{sect_inpainting}
\subsection{The missing data problem}
The problem of missing data can be formalized as:
\begin{eqnarray}
Y(t) = M(t) X(t),
\end{eqnarray}
with $X(t)$ the ideal complete (i.e. regularly sampled) time series, $Y(t)$ the observed time series (with gaps), and $M(t)$ the binary mask (i.e., window function with $M(t) = 1$, if we have information at data point $X(t)$; $M(t) = 0$ otherwise).
In Fourier space, the multiplication by the mask becomes a convolution:
\begin{eqnarray}
\hat{Y}(t) = \hat{M}(t) * \hat{X}(t),
\end{eqnarray}
where $\hat{}$ denotes the Fourier transform.
The convolution by the spectral window $\hat{M}(t)$ causes the energy at each frequency of the power spectrum to leak into surrounding frequencies, producing a spectral leakage in the Fourier domain. 

\subsection{Sparse {\it inpainting}}
\label{sect_inpainting_old}

In Paper I, we proposed to use a method of sparse {\it inpainting} 
to estimate the missing data.
The method proposed was introduced by \cite{inpainting:elad05} and consists in recovering $X(t)$ knowing $Y(t)$ and $M(t)$ by imposing a prior of sparsity on the solution $X(t)$.
This {\it inpainting} method already had some major successes in astrophysics (e.g. Weak Lensing \cite{inpainting:pires09, inpainting:jullo14}, CMB \cite{inpainting:rassat14}, Asteroseismology, \cite{inpainting:garcia14, inpainting:pires15}).
The sparse {\it inpainting} method uses the prior that there is a representation $\Phi^T$ of the time series $X(t)$ where most coefficients $\alpha = \Phi^T X$ are close to zero ($^T$ represents the transpose matrix). 
For example, if the time series $X(t)$ was a single sine wave, the representation $\Phi^T$ would be the Fourier transform because all but one coefficient of the Fourier representation of a sine are equal to zero.\\
The solution of this problem is obtained by solving:
\begin{equation}
\min  \| \alpha \|_1    \textrm{ subject to }  \parallel Y - MX   \parallel^2 \le \sigma^2,
\label{functional}
\end{equation}
where $||.||_1$ is the convex $l_1$ norm (i.e. $ || z ||_1 = \sum_k | z_k |$), $|| . ||$ is the classical $l_2$ norm (i.e. $|| z ||^2 =\sum_k (z_k)^2$) and $\sigma$ is the standard deviation of the noise in the observed time series.

The solution of such an optimization task can be obtained through an iterative algorithm introduced by \cite{inpainting:elad05}.
Let $X_i$ denotes the reconstructed time series at iteration $i$. If the time series is sparse enough in the representation $\Phi^T$, in this representation the largest coefficients should originate from the time series we want to recover. Thus, the algorithm is based on a threshold that decreases exponentially (at each iteration) from a maximum value to zero. By accumulating more and more high coefficients through each iteration, the gaps in $X_i$ are filling up steadily and the power of the coefficients due to the gaps is decreasing. This algorithm needs as inputs the observed incomplete data $Y$ and the binary mask $M$.\\

The algorithm can be described as follows:
\begin{enumerate}
  \item Set the maximum number of iterations $I_{max}=100$, the solution $X^0$ is initialized to zero, 
    the maximum threshold $\lambda_{max} = \max(\mid \Phi^T Y \mid)$ with $\Phi^T$ being a global Discrete Cosine Transform (DCT), and the minimum threshold $\lambda_{min} = 0$.
    \item Set $i = 0$, $\lambda^0 = \lambda_{max}$. Iterate:
    \item Set $U^i = X^i + M(Y-X^i)$ to enforce the time series to be equal to the observed data where the mask $M$ is equal to $1$.
    \item Compute the forward transform of $U^i$: $\alpha = \Phi^TU^i$.
    \item Compute the threshold level $\lambda^i = F(i,\lambda_{max},\lambda_{min})$, where $F$ is a function that describes the decreasing law of the threshold.
    \item Compute $\tilde \alpha$ by keeping only the coefficients $\alpha$ above the threshold $\lambda^i$ and setting the others to zero.
    \item Reconstruct $X^{i+1}$ from the remaining coefficients $\tilde \alpha$ : $X^{i+1} = \Phi \tilde\alpha$.
    \item Set $i=i+1$. If $i<I_{max}$, return to step 3.
\end{enumerate}
In Paper I, $\Phi^T$ is chosen to be a global DCT because it provides a sparse representation for the EPV signal. 
The function $F$ used to describe the threshold decreases (at each iteration $i$) from $\lambda_{max}$ to zero following the empirical law below: 
\begin{eqnarray}
F(i, \lambda_{\max}) = \lambda_{\max}  \left(1-\erf \left(\frac{i\beta}{N-1}\right) \right),
\end{eqnarray}
with $\beta = 2.8$. 
This law is commonly used because it follows the fast (i.e. exponential) decay of the coefficients that is commonly observed in a sparse representation.

\begin{figure}
\includegraphics[width=0.45\textwidth]{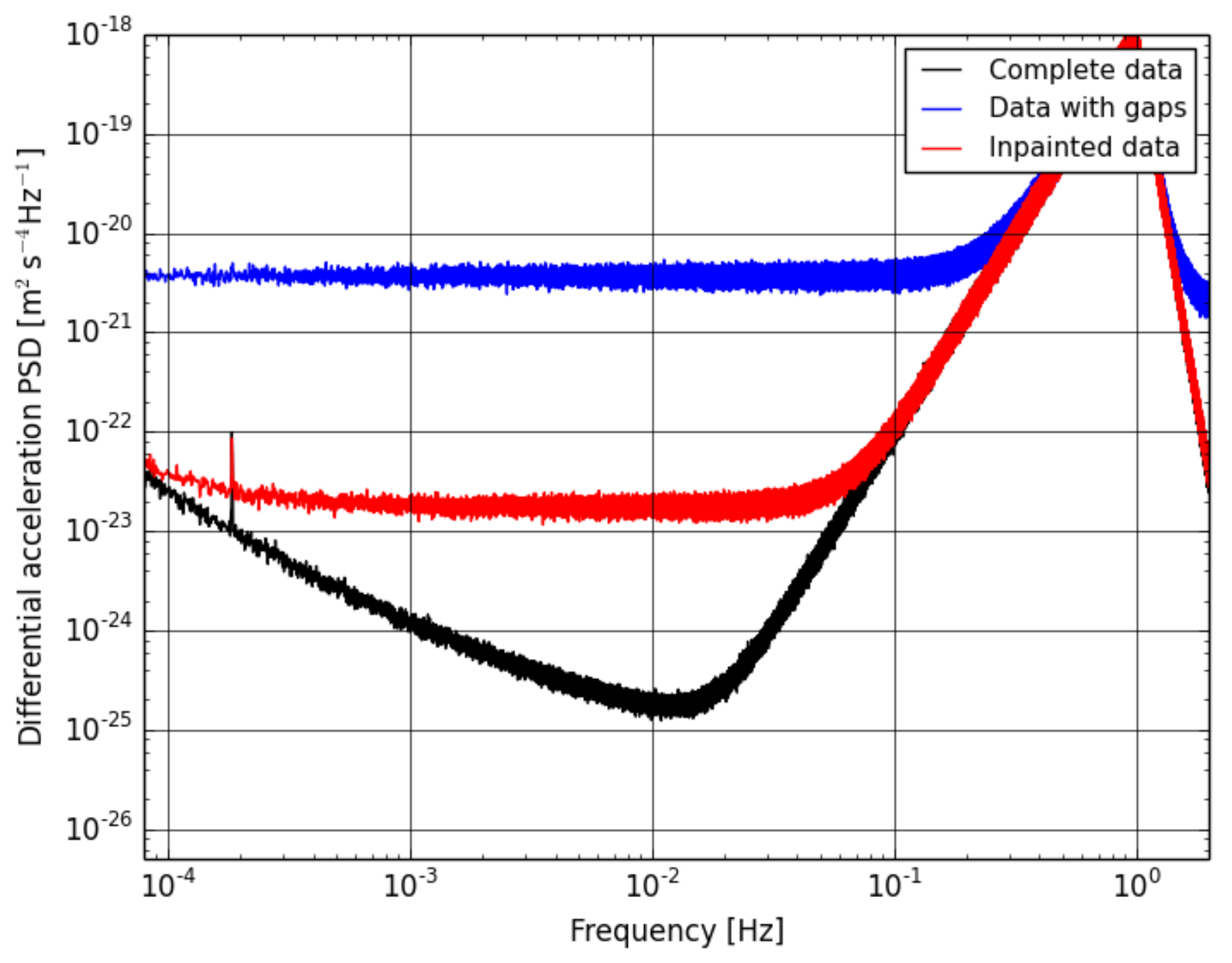}
\includegraphics[width=0.45\textwidth]{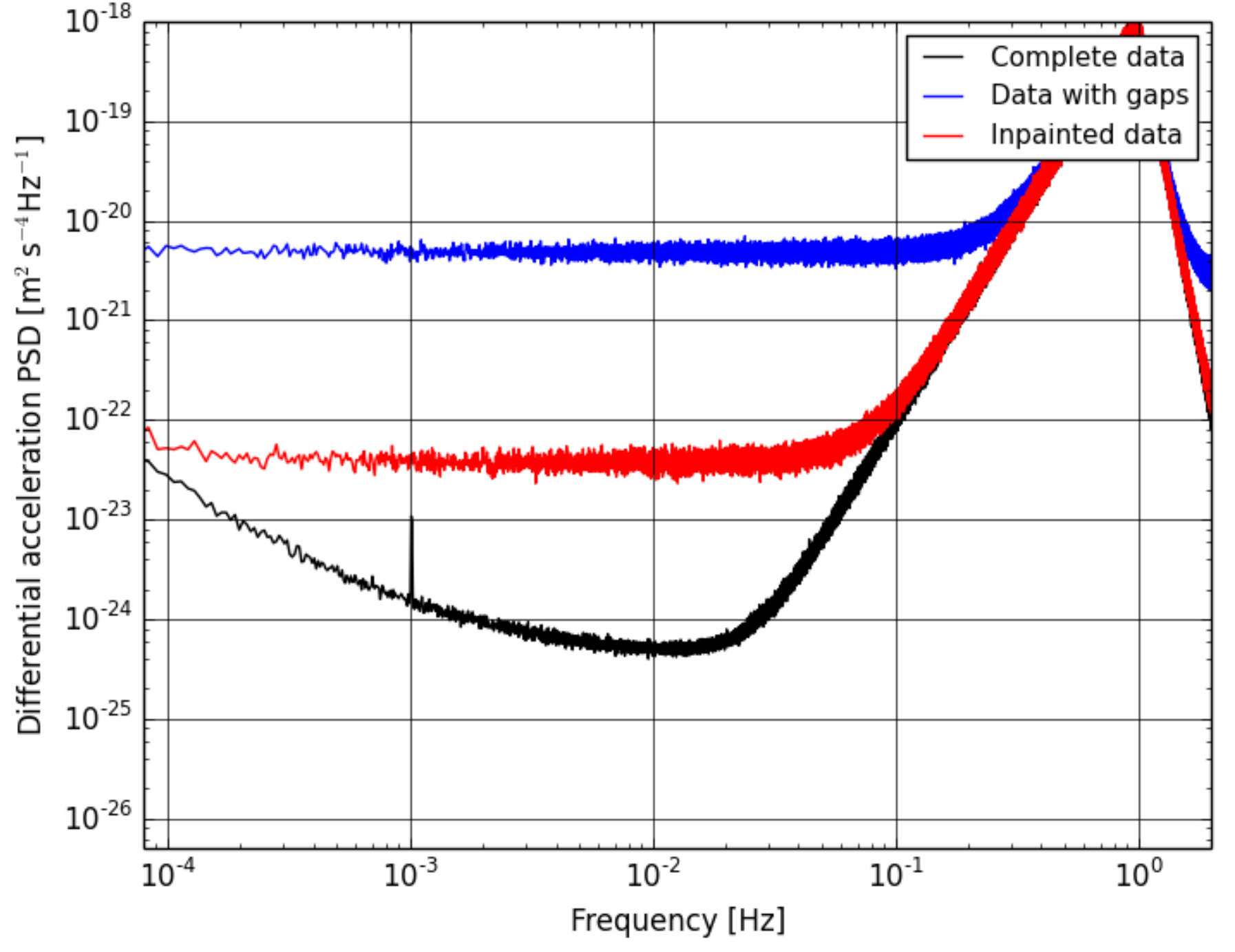}
\caption{Estimated MICROSCOPE differential acceleration power spectral density (PSD) averaged over 100 120-orbit simulations in the inertial mode (upper panel) and over 100 20-orbit simulations in the spin mode (lower panel). The black line shows the PSD when all the data is available, the blue line shows the effect of missing values and the red line shows the PSD estimated from data filled with the {\it inpainting} method developed in Paper I. Note the EPV peaks of $3 \times 10^{-15}$ at 1.8 $10^{-4}$ Hz (in the upper panel) and $10^{-3}$ Hz (in the lower panel).}
\label{fig_inpainting}       
\end{figure}

In Fig.~\ref{fig_inpainting}, the black curve shows the estimated MICROSCOPE differential acceleration PSD in the inertial mode averaged over 100 simulations of 120 orbits. The blue curve shows the effect of missing 3\% of the data. The red curve shows the PSD estimated after filling missing values using the {\it inpainting} algorithm just described 
above. 

In the inertial mode (upper panel), a least-squares fit of a sinusoidal function at $f_{\rm EP} = 1.8 \times 10^{-4}$ Hz allowed us to detect the EPV peak with a precision of $1.18\times10^{-15}$, which is very close to the MICROSCOPE requirements. 
However, if we wanted 
to confirm the detection of the EPV signal in the spin mode (lower panel), there is still an important spectral leakage in the intermediate frequency region where the EPV peak is expected for this second configuration ($f_{\rm EP} = 10^{-3}$ Hz). Note also that the peak in the spin mode appears smaller in the PSD representation due to the different integration time.

\subsection{Sparse {\it inpainting} improvements: ICON}
\label{sect_inpainting_new}

The {\it inpainting} developed in Paper I gives really promising results in the inertial mode. However, it remains unsatisfactory to detect and characterize a possible EPV peak in the spin mode. 
Fig.~\ref{fig_inpainting} shows that there is still a residual spectral leakage from the high-frequency noise peak to the low frequency region of the spectrum where we try to detect a possible EPV signal.
In this section, we describe the improvements we brought to the algorithm.

\subsubsection{Noise constraint}
\label{noise_section}
A major problem in the previous version of the inpainting, is that the minimisation problem of eq.~\ref{functional} is only optimal for white noise data. However, this minimization can be extended to the colored-noise case. An effective way to do so and thus reduce the spectral leakage is to introduce a prior on the noise.
This prior could be easily introduced in the algorithm 
by forcing, at each iteration, the high frequency part of the spectrum to follow an a priori model of the noise. However, this would break the required model-independence of the method.
To bypass this problem, so that the method remains model-independent, the noise constraint is based on the data.

To do so, at each iteration $i$ of the algorithm, we perform a wavelet transform of the signal $X^i(t)$ at this iteration using the {\it {\`a} trous} algorithm:
\begin{equation}
X^i(t) = {c_{J}}(t)+ \sum_{l=1}^{J} w_{l}(t),
\end{equation}
where $J$ is an input parameter, $c_{J}$ is a smooth version of the original signal $X^i(t)$ and $w_l$ are the wavelet bands that give the details of the signal $X^i(t)$ at different resolutions (see Starck et al.~\cite*{book:starck02, book:starck06} for details).  
Thus, if the signal $X^i(t)$ is of size $N$, the algorithm outputs $J+1$ arrays of size $N$. 
In this application, J is chosen to be equal to 10 in the spin case and 14 in the inertial case to properly handle the residual spectral leakage at the position of the EPV peak.

The wavelet filters $\psi(x)$ used for this application are defined as
\begin{equation}
\frac{1}{2} \psi(\frac{x}{2} )=\varphi(x) - \frac{1}{2}  \varphi(\frac{x}{2}),
\end{equation}
 the difference between two B$^3$-Spline functions $\varphi(x)$ at two different resolutions with:
\begin{equation}
\varphi(x)=\frac{1}{12}(|x-2|^3-4|x-1|^3+6|x|^3-4|x+1|^3+|x+2|^3).
\end{equation}

\begin{figure}
\includegraphics[width=0.4\textwidth, height=0.2\textheight]{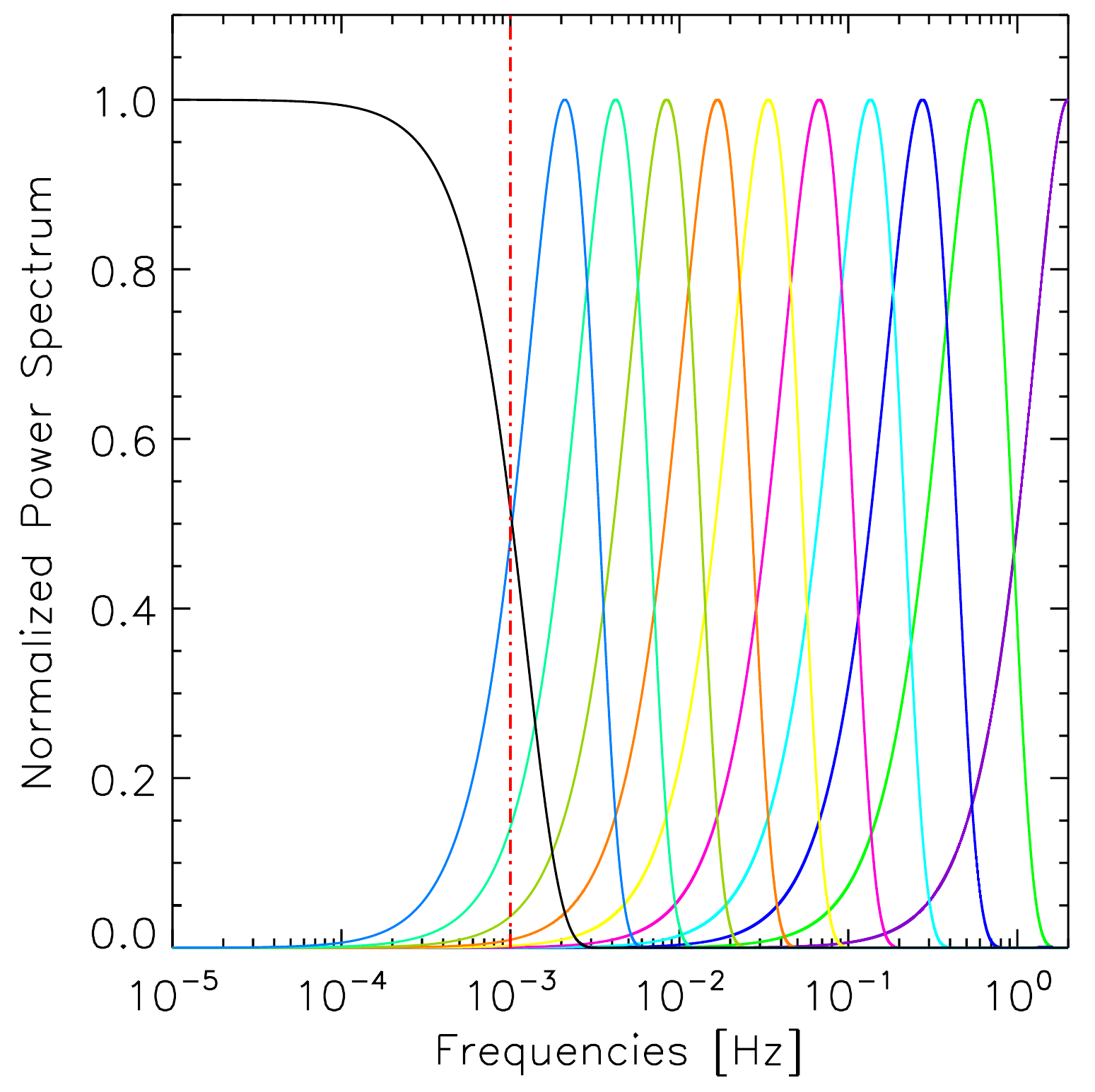}
\caption{Frequency response of the wavelet filters (in different colors) and smoothing filter $\varphi(x)$ (in black). The wavelet filters have been used to add a constraint about the power spectrum of the noisy signal. Each wavelet band $w_l$ is obtained by the convolution of the signal with
the wavelet filter functions of various characteristic scales as described in the text}. Thus, each wavelet band contains the frequency information of the band selected by the corresponding wavelet filter.
\label{fig_wavefilters}       
\end{figure}

Thus, at a given scale $l$, the wavelet filters are defined as
\begin{equation}
\frac{1}{2^{l+1}} \psi(\frac{x}{2^{l+1}} )=\frac{1}{2^{l}}  \varphi(\frac{x}{2^{l}}) - \frac{1}{2^{l+1}}  \varphi(\frac{x}{2^{l+1}}).
\end{equation}
Fig.~\ref{fig_wavefilters} shows the shape of the power spectra of the wavelet filters for $l = 0$ to $10$ (in colors) and the smoothing filter $\varphi(x)$ (in black) used for the spin case.
Note the position of the EPV peak at $f_{\rm EP} = 10^{-3}$ Hz, represented in the figure as a vertical dashed red line. The wavelet filters derived in this way have a compact support in real space and are well localized in the Fourier domain.
 Additionally, the wavelet decomposition is very fast.
 
Once the signal is decomposed into several wavelet bands $l$, the estimation of the standard deviation of $w_l(t)$  with $t$ constrained to be outside gaps enables us to estimate the mean power spectrum of the noisy signal in this frequency band. 
Hence, we have now a way to estimate a broad-band power spectrum of the noisy signal from incomplete data.
This being said, 
for each wavelet band $l$ the code finds the standard deviation of $w_l(t)$ with $t$ constrained to be outside gaps, and does the same for $t$ constrained to be inside gaps. 
And then, $w_l(t)$ is rescaled inside the gaps by a constant $\nu_l$ chosen so that the standard deviation inside gaps is the same as outside gaps for that $l$.
Thus, we reduce the spectral leakage by imposing that constraint for each wavelet band.
In this process, the f$_{\rm EP}$ frequency is considered to avoid a rescaling of a possible EPV signal.
This is the major improvement of the code.

\subsubsection{Threshold law}

The way the threshold is decreased at each iteration has also an impact on the results.
Ideally, we would like to have a number of iterations as large as the number of points in the time series and decrease the threshold in such a way so to 
add one single additional coefficient to $\alpha$ at each iteration.
However, this number is too large and instead, we have to find a trade-off between the speed of the algorithm and its quality. 

We optimize the decreasing law $F$ for the MICROSCOPE data by modifying slightly the slope ($\beta = 4.8$), and we add a constant $\rho$ corresponding to the ratio of missing values:
\begin{eqnarray}
F(i, \lambda_{max}) = \rho \lambda_{max}  \left(1-\erf \left(\frac{i\beta}{N-1}\right) \right).
\end{eqnarray}
The number of iterations $I_{max}$ has been raised to 1000. Higher values for $I_{max}$ have a small impact on the result. The value $\rho$ has been chosen equal to $0.03$ in the inertial mode (resp. 0.04 in the spin mode) which is the ratio of missing values. This value ensures that the maximum threshold $\lambda_{max}$ will be larger than the ``plateau" due to the spectral leakage before {\it inpainting} (see the blue curve in Fig.~\ref{fig_inpainting}).
Indeed, the algorithm does not need to spend time on high-power density values of the threshold where the spectral leakage is negligible.
Thus, the constant $\rho$ makes the threshold start at a lower value and the new value for the slope of the decreasing law $\beta$ makes the threshold decrease more quickly to the low amplitudes where the EPV signal is expected. With this new decreasing law, most of the coefficients due to the high-frequency noise are caught at the first iterations, thus saving more iterations for the small coefficients where the spectral leakage is high. 

\subsubsection{Algorithm}

The new algorithm can be described as follows:

\begin{enumerate}
  \item Set the maximum number of iterations $I_{max}=1000$, the solution $X^0$ is initialized to zero, 
    the maximum threshold $\lambda_{max} = \max(\mid \Phi^T Y \mid)$ with $\Phi^T$ a global Discrete Cosine Transform (DCT) and the minimum threshold $\lambda_{min} = 0$.
    \item Set $i = 0$, $\lambda^0 = \lambda_{max}$. Iterate:
    \item Set $U^i = X^i + M(Y-X^i)$ to enforce the time series to be equal to the observed data where the mask $M$ is equal to $1$.
    \item Compute the forward transform of $U^i$: $\alpha = \Phi^TU^i$.
    \item Compute the new threshold level $\lambda^i $:\\ $\lambda^i =  \rho \lambda_{max}  \left(1-\erf \left(\frac{i\beta}{N-1}\right) \right)$  \\
    with $\beta = 4.8$.
    \item Compute $\tilde \alpha$ by keeping only the coefficients $\alpha$ above the threshold $\lambda^i$ and setting the others to zero.
    \item Reconstruct $X^{i+1}$ from the remaining coefficients $\tilde \alpha$ : $X^{i+1} = \Phi \tilde\alpha$.
    \item Set $X^{i+1}(t')= {c_{J}}(t')+  \sum_{l=1}^{J} \nu_l  w_{l}(t')$ for $t'$ inside the gaps to apply the noise constraint described in section \ref{noise_section}.
    \item Locate the position of the EPV signal in the DCT and remove the effect of the noise constraint at the position of the peak.
    \item Set $i=i+1$. If $i<I_{max}$, return to step 3.
\end{enumerate}

\section{Results} 
\label{sect_results}

\subsection{Simulations}
\label{sect_simulations}
To assess {\it inpainting}'s performance on MICROSCOPE-like data, we design a suite of simulations with the assumption that all nuisance parameters are perfectly corrected for.
Hence, the signal consists of just a pure sine at a well known frequency and a noise: 
\begin{eqnarray}
y_{\rm EP}(t) = \delta g_{\rm EP}(t)/2 + \mathcal{N}(t).
\end{eqnarray}
The simulated time series are sampled at $f_s = 4 \text{ Hz}$.
For the sake of clarity in this section, and to have an acceptable signal-to-noise ratio, we set $\delta=3\times10^{-15}$ following \cite{karma:baghi15} and Paper I.\\

We then consider the two satellite configurations described in Sect~\ref{sect_microscope}:
\begin{itemize}
\item Inertial mode: 120 orbits, $f_{\rm EP, iner} =1.8\times10^{-4}$ Hz
\item Spin mode: 20 orbits, $f_{\rm EP, spin} = 10^{-3}$ Hz
\end{itemize}

Following the experimental setup used in Paper I, we define missing values in a worst case scenario, with 3\% of missing values (resp. 4\% of missing values) in the inertial mode (resp. in the spin mode), due to 260 tank crackles per orbit, 24 (resp. 111) MLI coating crackles per orbit in the inertial mode (resp. spin mode), 0.2 micrometeorite impacts per orbit and 0.05 telemetry loss per orbit. The mean duration of saturated data due to crackles and micrometeorite impacts is set to 0.75 seconds (corresponding to 3 data points), and the telemetry losses can vary from 1 second to 250 seconds. Gaps are distributed randomly within the time series, following a uniform distribution.
Gaps are not pre-defined, but their distribution is drawn randomly for each simulation, therefore we have access to their statistics only. The exact probability of occurrence of these events is unknown at the time of writing. However, the worst case scenario used in the simulations have been estimated by on-ground tests.
Finally, we generate 100 similar simulations for each configuration to perform a statistical analysis of our estimates.\\

\subsection{Missing data interpolation}
\begin{figure}
\includegraphics[width=0.45\textwidth]{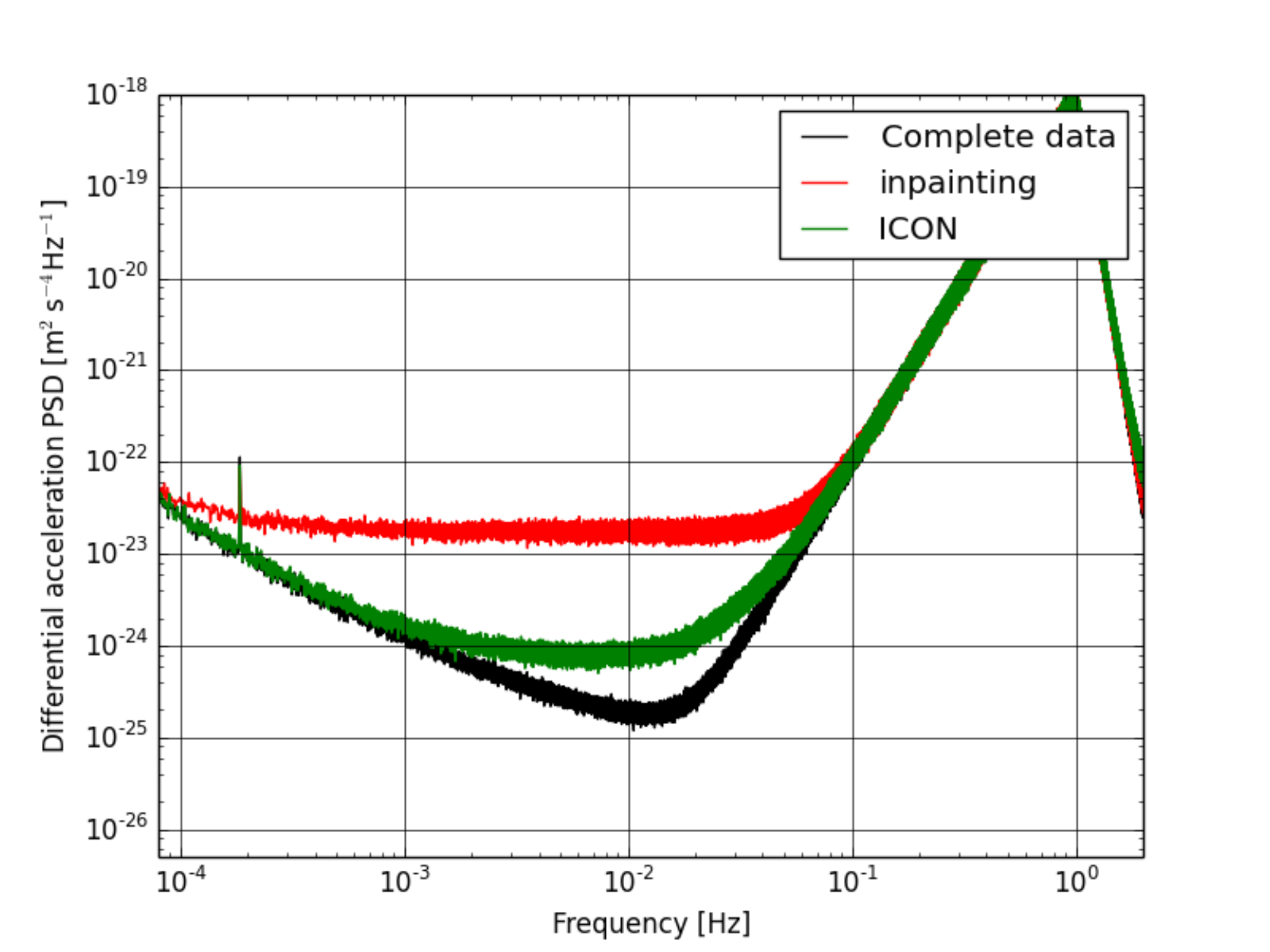}
\includegraphics[width=0.45\textwidth]{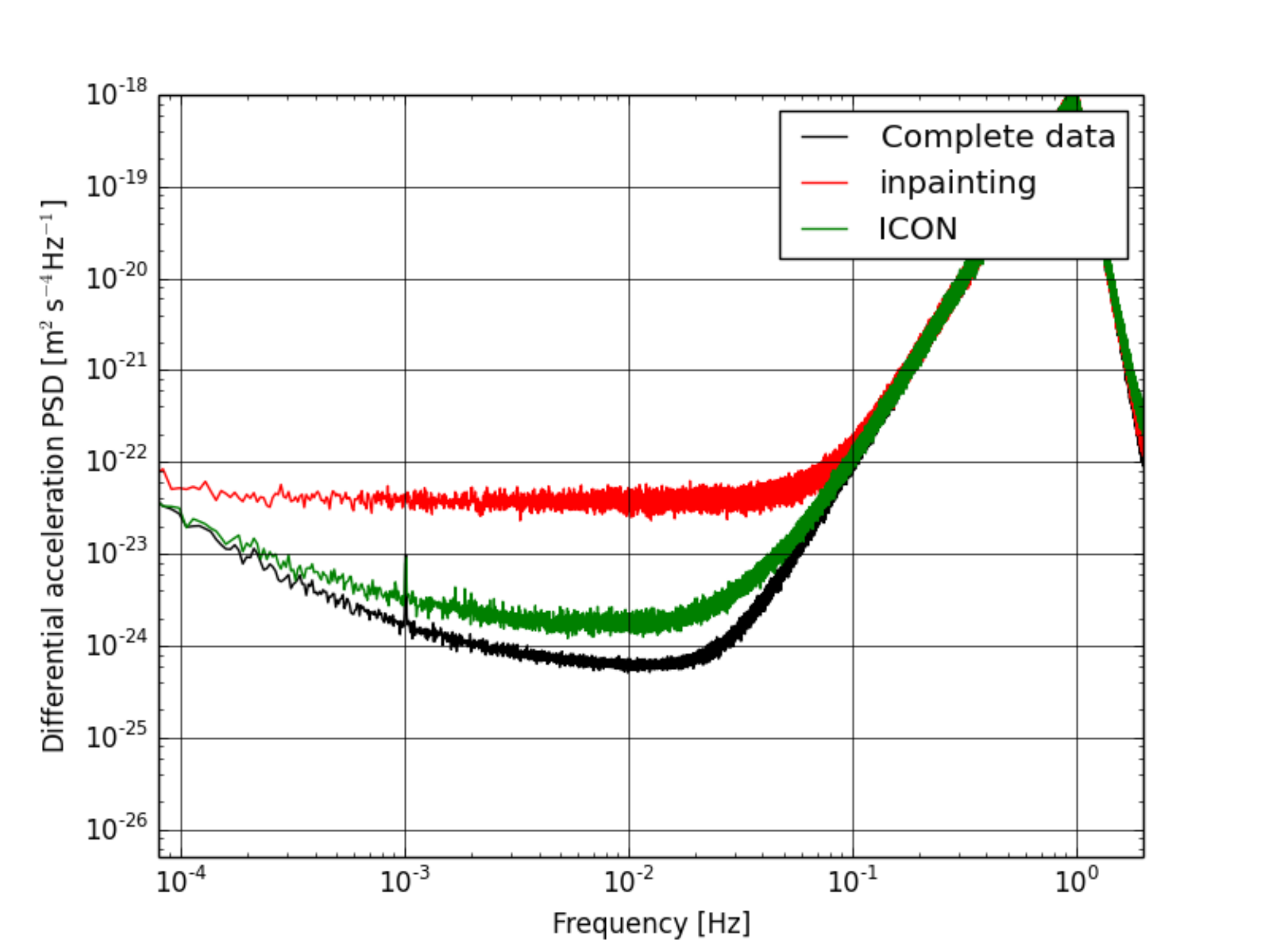}
\caption{
MICROSCOPE differential acceleration PSD estimates averaged over 100 simulations in the inertial mode (upper panel) and in the spin mode (lower panel). The black lines show the PSD estimated when all the data is available, the red lines show the PSD estimated from data filled with the {\it inpainting} method developed in Paper I and the green lines show the PSD estimated from data filled with the new {\it inpainting} method (ICON) presented in this paper.}
\label{fig_inpainting_new}       
\end{figure}

The sets of simulations presented above are analyzed using the {\it inpainting} method developed in Paper I (described in Sect.~\ref{sect_inpainting_old}) and the new version of the code, ICON presented in this paper (described in Sect.~\ref{sect_inpainting_new}).
The PSD estimates averaged over 100 simulations obtained with these two methods are shown in Fig.~\ref{fig_inpainting_new} for the inertial mode (upper panel) and for the spin mode (lower panel). 
The black curves show the averaged estimated MICROSCOPE differential acceleration.  
The red curves show the averaged PSD after filling missing values using the {\it inpainting} method developed in Paper I and the green curves show the averaged PSD after filling values with the new algorithm proposed in this paper.

It is obvious that the original PSD is better recovered with the {\it inpainting} method presented in this paper. Indeed, the spectral leakage in the PSD estimate is more than one order of magnitude smaller than the residual spectral leakage obtained with the {\it inpainting} method developed in Paper I which already reduces the spectral leakage by more than two order of magnitudes if compared to the PSD estimated from incomplete data (see Fig. \ref{fig_inpainting}). Although the residual spectral leakage has not totally disappeared, the new version of the {\it inpainting} algorithm allows the EPV signal to clearly emerge from the noise both in the inertial and spin modes. 
It is therefore possible to detect and estimate the amplitude of the signal in both configurations after running the improved algorithm.

\subsection{Detection and characterization of the EPV }

To further quantify the results obtained with {\it inpainting}, we can use common techniques to detect and characterize the EPV signal since the {\it inpainting} interpolation enables us to recover a regularly sampled time series.

In this work, because we are assuming perfect correction of systematics and instrumental's calibration, the signal we are looking for is just a pure sine wave, of known frequency and phase. 
Thus, a simple least-squares fit to the corrected data is sufficient to estimate the amplitude of the EP parameter $\delta$. For more realistic signals, a more general regression technique will be required, like a least-squares fit with a more complex model or a MCMC technique, which will allow us to constrain more parameters.




As aforementioned, we have run 100 realizations for the two configurations described in Sect.~\ref{sect_microscope} to perform a statistical analysis of our estimate.
The errors we quote are the rms of the least-squares estimators estimated on these 100 simulations; in this way, we are able to quantify the combination of errors coming both from the {\it inpainting} interpolation and from the least-squares estimation. 

In this work, because we have noted a bias in the amplitude of the EPV peak estimated after inpainting, we adopt an ordinary least-squares estimator different from the Paper I. We perform a simple least-squares fit in the temporal domain to remove all possible bias introduced by the method.
The results of the least-squares estimation are summarized in Table~\ref{table_results}.\\

\begin{table*}[htb]
\begin{center}
\begin{tabular}{|c||c|c||c|c|}
\hline
& \multicolumn{2}{c||}{inertial configuration} & \multicolumn{2}{|c|}{spin configuration}\\
\hline
\hline
&$<\hat\delta>$ & $\sigma_{\hat\delta}$ &$<\hat\delta>$ & $\sigma_{\hat\delta}$ \\
\hline
\hline
Complete data   & $3.15\times10^{-15}$  & $0.90\times10^{-15}$& $2.97\times10^{-15}$  & $0.60\times10^{-15}$\\
\hline
Incomplete data & $4.63\times10^{-15}$  & $14.66\times10^{-15}$& $0.61\times10^{-15}$  & $39.01\times10^{-15}$\\
\hline
Inpainted data (Paper I) & $2.40\times10^{-15}$  & $1.18\times10^{-15}$& $2.53\times10^{-15}$  & $3.08\times10^{-15}$\\
\hline
Inpainted data (ICON) & $2.74\times10^{-15}$  & $0.96\times10^{-15}$& $2.05\times10^{-15}$  & $1.20\times10^{-15}$\\
\hline
\end{tabular}
\end{center}
\caption{EPV signal estimation and statistical errors for a simulated EPV peak of $3 \times 10^{-15}$. These numbers correspond to the mean and the standard deviation of a time-domain least-squares estimators obtained on a set of 100 simulations.}
\label{table_results}
\end{table*}



Table~\ref{table_results} shows that while the inpainting algorithm of paper I strongly decreases the EPV measurement uncertainty $\sigma_{\delta}$, further improvement can be expected from the new (ICON) method presented in this paper, especially for the spin mode.
Therefore, the new version of the {\it inpainting} code allows us to have a significant measurement of a $3\times10^{-15}$ EPV signal in both configurations, which would be impossible by simply performing an Ordinary Least Square fit on the available data. Given our estimated $1\sigma$ statistical error, we can conclude that with only one measurement run of 120 orbits (inertial mode) or 20 orbits (spin mode), we may be able to characterize a possible EPV signal with a $0.96\times10^{-15}$ precision in the inertial mode and $1.20\times10^{-15}$ in the spin mode assuming an instrumental noise at the level of the simulated one.

However, in view of the number of simulations ($N = 100$) used to perform the statistical analysis, the results show a bias on the estimated mean value of the EPV peak in the spin mode. Indeed, the expected standard deviation of the mean should be $\sigma_{\rm mean}=\sigma/\sqrt{N}$ = $0.12 \times 10^{-15}$ in the spin mode (resp. $0.096 \times 10^{-15}$  in the inertial mode) and the estimated mean value of the EPV peak is outside the $3\sigma_{\rm mean}$ error bars in the spin mode. \\

\begin{figure}
\includegraphics[width=0.5\textwidth]{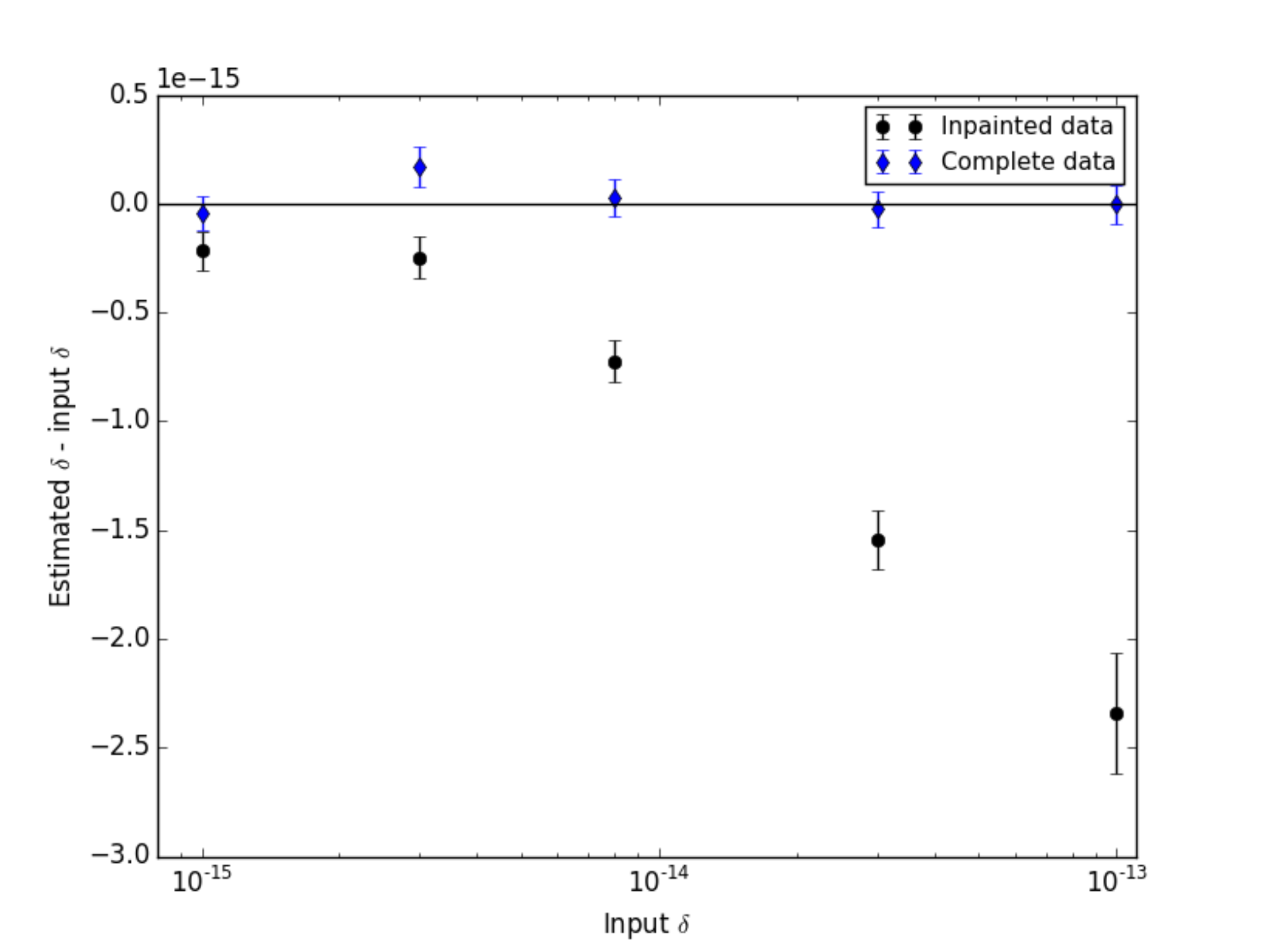}
\includegraphics[width=0.5\textwidth]{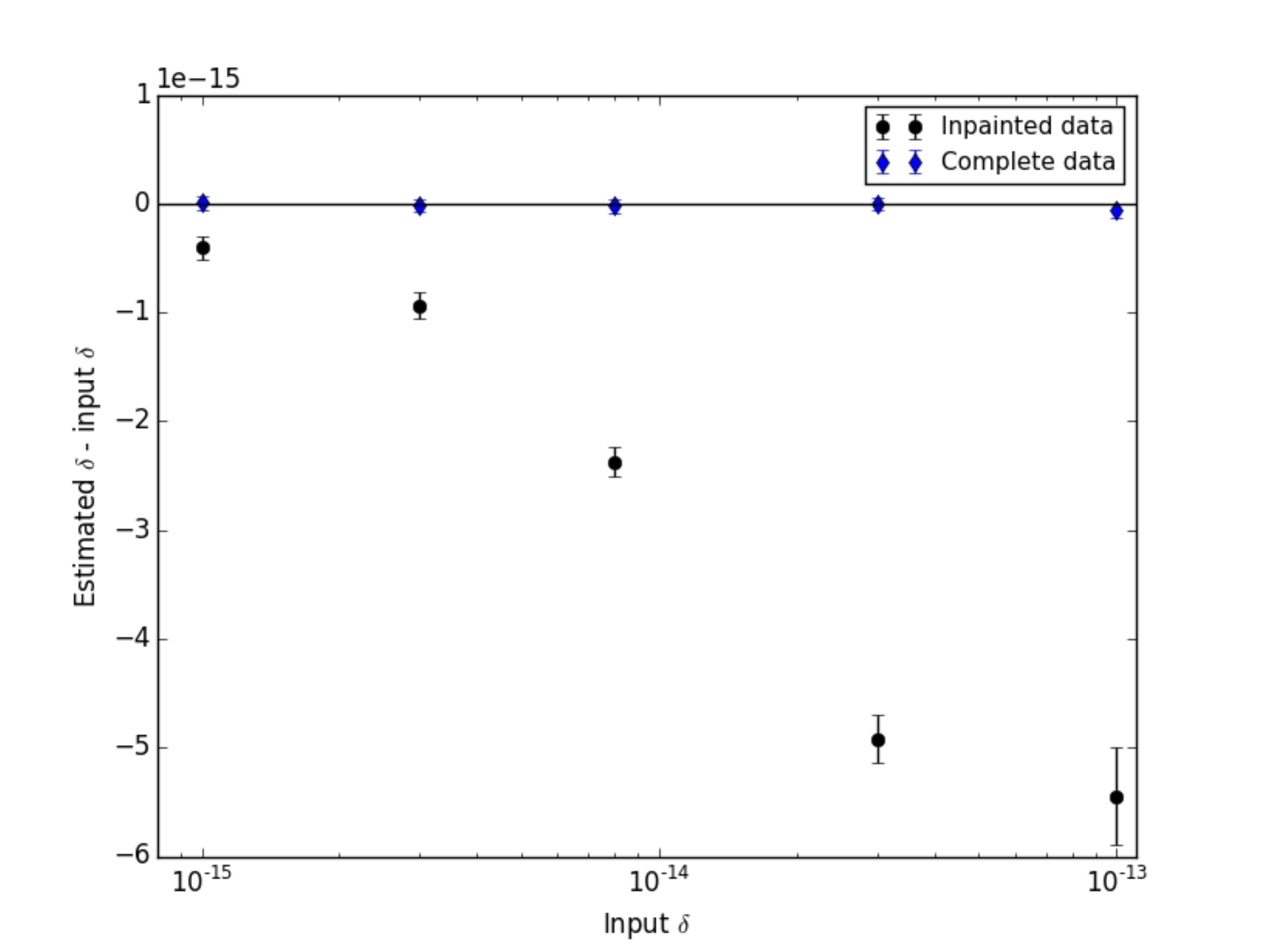}
\caption{EPV signal estimation for different amplitudes ($10^{-15}$, $3 \times 10^{-15}$, $8 \times 10^{-15}$, $3 \times 10^{-14}$, $10^{-13}$) in the inertial case (upper panel) and in the spin case (lower panel). The points correspond to the mean and the error bars to the standard deviation of the mean obtained on a set of 100 simulations.}
\label{fig_inpainting_ampli}       
\end{figure}

We further investigated the bias introduced by inpainting by considering different amplitudes for the EPV signal ($10^{-15}$, $3 \times 10^{-15}$, $8 \times 10^{-15}$, $3 \times 10^{-14}$, $10^{-13}$) in the two satellite configurations (see Fig.~\ref{fig_inpainting_ampli}).
The bias is more important in the spin mode. Indeed, in the inertial mode (upper panel), the bias appears for EPV peaks larger than $3 \times 10^{-15}$ and in the spin mode (lower panel), a bias is significant even for the small amplitudes of the peak.
The likely reason for the bias in the EPV signal estimation is that the sparsity condition in the inpainting method is not fully verified.
Although the decomposition into a set of oscillating functions of the DCT is ideal to represent the EP sine-wave signal ($\delta g_{\rm EP}(t)/2$), the global DCT is much less efficient to represent the continuous spectrum of the colored-noise. Indeed, we checked that the bias disappears in white-noise simulations.
Although the bias seems to increase with the amplitude of the EPV peak, the relative bias decreases with the amplitude. 
The bias is related to the noise spectral leakage: the more the peak is embedded in the noise after spectral leakage, the larger the bias. This explains why the bias is more important in the spin mode.


The {\it inpainting} code (ICON) presented in this paper is dedicated to reliably asserting the detection of a possible EPV signal in parallel with the other code present in the pipeline, KARMA \cite{karma:baghi15}. Having these two independent techniques allow us to cross-check our results. 
If an EPV peak is detected and confirmed, further work will be needed to fully characterize the peak.

\section{Conclusion} 
\label{sect_conclusion}

We have presented an updated version of the {\it inpainting} algorithm used to 
judiciously fill-in the missing values in MICROSCOPE data.
Several improvements have been made to lower the noise spectral leakage residuals of the {\it inpainting} method.
The major improvement was obtained by the introduction of a prior on the noise power spectrum. We showed how this prior can be  directly derived from the incomplete data using a multi-scale representation.
The second improvement consists in changing the threshold law of the iterative algorithm used to solve the minimization problem in Eq.~\ref{functional}. The idea behind this modification is to spend more time on small coefficients (depending on the missing data ratio) where the spectral leakage is more important.\\

The performance of the new {\it inpainting} code was assessed based on MICROSCOPE simulations in a worst-case scenario for missing data assuming perfect correction of systematics and perfect instrumental calibration.
Further work is under way  to test the code in more realistic simulated data including calibration imperfections and other additional perturbations (Berg\'e et al in prep). 
We showed that the performance of the new {\it inpainting} algorithm presented in this paper reach the MICROSCOPE requirements for both the inertial and spin modes. 
With the simulated noise, our estimated statistical $1\sigma$ error for the detection of a $3\times10^{-15}$ EPV signal is $0.96\times10^{-15}$ in the inertial mode and $1.20\times10^{-15}$ in the spin mode.
Thus, the new version of the {\it inpainting} code will replace the previous version in the official MICROSCOPE's data processing and analysis pipeline.\\

In the performance study, we noticed a bias in the estimation of the EPV signal that is explained by the fact that the sparsity constraints is not fully verified because the colored-noise is not sparse enough in the DCT representation.
This bias in the EPV signal estimation makes the {\it inpainting} code suboptimal to characterize a possible EPV signal with an ordinary least-squares method.
The major asset of the {\it inpainting} technique is that it is model-independent; this allows us to cross-check any EPV signal detection with the KARMA independent method. The characterization of the EPV signal, if any detection is confirmed will require further work.\\

While the {\it inpainting} method presented in this paper has been optimized to process MICROSCOPE simulated data, it also provides a robust method to deal with missing data in the general context of time series dominated by an unknown colored-noise. This is because the code is model-independent, fully adaptive to the data and should behave well in more complex data. Following the reproducible research guidelines, the {\it inpainting} software presented in this study, named ICON ({\bf I}npainting for {\bf CO}lored-{\bf N}oise dominated signals), is now freely available at the following address: \url{http://www.cosmostat.org/software/icon/}.\\

\begin{acknowledgements}
We thank Patrice Carle for his help in incorporating the inpainting code to the official MICROSCOPE data processing software. We wish also to thank Florent Sureau for useful discussions.
This work makes use of technical data from the CNES-ESA-ONERA-CNRS-OCA Microscope mission, and has received financial support from ONERA and CNES.
We acknowledge the financial support of the UnivEarthS Labex program at Sorbonne Paris Cit\'e (ANR-10-LABX-0023 and ANR-11-IDEX-0005-02).
\end{acknowledgements}

\bibliography{inpainting}
\end{document}